\documentclass[aps,prl,superscriptaddress,showpacs,preprintnumbers,amsmath,amssymb]{revtex4}

\usepackage{graphicx}
\usepackage{appendix}
\begin{document}
\title{Adiabatic theory of the polaron spectral function}
\author{V. V. Kabanov}
\affiliation {\it Josef Stefan Institute 1001, Ljubljana,
Slovenia}

\begin{abstract}
An analytic theory for the spectral function of electrons coupled with phonons is formulated in the adiabatic limit. In the case when the chemical
  potential is large and negative $\mu \to -\infty$ the ground state does not have the adiabatic deformation and the spectral function is defined by the
  standard perturbation theory. In this limit, we use the diagram technique in order to formulate an integral equation for the renormalized vertex. The spectral
  function was evaluated by solving Dyson's equation for the self-energy with the renormalized vertex. The moments of the spectral function satisfy the exact
  sum rules up to the 7th moment. In the case when the chemical potential is pinned at the polaron binding energy the spectral function is defined by the ground
  state with a nonzero adiabatic deformation. We calculate the spectral function with the finite polaron density in the adiabatic limit. We also demonstrate how the
  sum rules for higher moments may be evaluated in the adiabatic limit. Contrary to the case of zero polaron density the spectral function with the finite polaron
  concentration has some contributions which are characteristic for polarons.

\end{abstract}
\maketitle
\section{Introduction}
The properties of different types of polarons were well studied a long time ago.
A number of review articles (see for example \cite{AustinMott,Elliott,AlexandrovMott,AlexandrovKrebs,Mishchenko,DevreeseAlexandrov,Klinger1})
and textbooks \cite{Pekar,KuperWhitfield,Appel,AlexandrovDevreese,AlexandrovMott2,Emin}
is available, which describe the spectroscopic, thermodynamic, kinetic, and other physical properties of polarons.
Usually, the polaron theory is used in order to describe electric transport in low mobility crystalline or organic semiconductors (see review article of
I. G. Austin, N. F. Mott \cite{AustinMott} and references therein or more recent papers on organic semiconductors \cite{Coropceanu}).
It was also successfully used in order to describe equilibrium and photo-induced mid-infrared optical absorption spectra of high-T$_c$ superconductors at low
doping \cite{Zamboni,Mihailovic,Falk,Calvani}. The description is based on the well developed theory of polaron optical absorption
\cite{Eagles,Klinger2,Reik,Emin1993,AKR_physicaC}.
The small Jahn-Teller polarons \cite{Hock} are also found in colossal magnetoresistance manganites \cite{Tokura,Millis}.

Permanent interest in polaron physics is not only related to the fact that polarons are found in many advanced functional materials.
Interest in this field has recently gone through a vigorous revival because
the polaron theory represents a very interesting testing ground for different numerical
techniques such as exact diagonalization calculations \cite{RanningerThibblin,AlexandrovKabanovRay,Bonca1,Wellein,Capone,Alvermann2010}, Quantum Monte Carlo simulations
\cite{Raedt,Pasha1,Pasha2,Mishchenko2}, and others \cite{Romero,Jeckelmann,Berciu,Paganelli2006}.

Note that until recently most investigations were related to the ground state properties of polaronic systems, such as polaron binding energy, effective mass, transport
properties as well as mid-infrared absorption. The development of angle resolved photoemission spectroscopy (ARPES) demands the analysis of the whole spectral function of
polaron systems \cite{AlexandrovRanninger}. This problem was successfully addressed by a number of different numerical techniques
\cite{RanningerThibblin,AlexandrovKabanovRay,Bonca1,Berciu,Hohenadler2003,Filippis2005,Cataudella2008,Vidmar2010,Goodvin2010,Bonca2}. Nevertheless, most of these
techniques are restricted to studying the so called Holstein model \cite{Holstein} of molecular crystals, where both the electron-phonon coupling constant and the phonon
frequency is momentum independent. The spatial dispersion of the electron-phonon interaction leads to a substantial increase in the number of nonzero matrix elements in the
Hamiltonian and poses substantial restrictions for this type of calculation. For the same reason, calculations are usually performed in the non-realistic limit for crystalline
materials when the phonon frequency $\omega_0$ is of the order of the hopping integral $t$ for electrons
\cite{AlexandrovKabanovRay,Wellein,Capone,Bonca1,Hohenadler2003,Filippis2005,Bonca2}. Moreover, the calculations are usually restricted by a 1D finite system. The system
size is usually restricted by $N\sim 10$ \cite{Bonca2} leading to poorly controllable finite size effects. On the other hand, it is well known that the polaron formation
crucially depends on the dimension of the system \cite{Emin,Kabanov}.

The investigation of the polaron spectral function was stimulated when the exact sum rules for dilute electron-phonon systems were formulated by Kornilovitch \cite{Pasha3}
(see also Ref.\cite{Berciu}). It was shown that the spectral function proposed in Ref.\cite{AlexandrovRanninger} satisfies only the zero order sum rule. On the other hand,
the numerical calculations in Ref.\cite{Bonca2} obey the sum rules derived for the limit of zero polaron density $n=0$. In Ref.\cite{Berciu} an uncontrollable approximation
of the polaron Greens function which neglects all momentum dependence of the self-energy was proposed.

Here we present an adiabatic theory for the spectral function of the Holstein model. The theory is based on equations, formulated for the case of the Holstein model
\cite{Holstein} in Ref.\cite{Kabanov}. These equations are similar to that, derived by Pekar for the polaron in polar crystals \cite{Pekar}. The equations have two different
sets of solutions \cite{Kabanov}. Near the trivial solution in the limit of zero polaron density $n=0$ (chemical potential $\mu\to-\infty$) we derive an equation for the
self-energy. Vertex in this equation is found from an equation that accurately takes into account threshold effects. This theory corresponds to the summation of infinite
series of diagrams where each phonon line has not more than three crossings. The theory takes into account exactly all diagrams up to 6th order.  As a result, the spectral
function obeys the sum rules up to the 7th order and has accuracy better than 3\% for $\lambda \le 3$ in the adiabatic limit. Here $\lambda$ is dimensionless electron-phonon
coupling constant. It is interesting to note that the polaron contribution to the spectral function is at least exponentially small $\sim \exp{(-const\sqrt{M/m})}$ where $m$
is the effective mass of the electron and $M$ is the ionic mass. As it was mentioned in Ref.\cite{Berciu} these small terms are absent in the sum rules for the spectral function
at zero polaron density, indicating that contributions from the polaron state to the spectral function are negligible. For the case when the chemical potential is pinned to
the polaron binding energy and small but finite concentration of polarons $n\neq0$, the ground state has nonzero adiabatic deformation. In this limit, we derive the spectral
function which obeys the exact known sum rules for finite polaron density.

The paper is organized as follows. In the next section, we briefly discuss some important details of the polaron theory, and then in the section "Results"
we discuss the polaron spectral function and the sum rules in the limit of zero and finite polaron density.

\section{Polaron theory}
\subsection{Electron-phonon interaction Hamiltonian}

The Hamiltonian of interacting electrons and phonons has the form:
\begin{eqnarray}
H &=& \sum_{\mathbf k}\epsilon(\mathbf k)c^{\dag}_{\mathbf k}c_{\mathbf k}+\frac{1}{\sqrt{N}}\sum_{\mathbf k,\mathbf q}\gamma(\mathbf q)\omega_{\mathbf q}
(c^{\dag}_{\mathbf k}c_{\mathbf k-\mathbf q}b_{\mathbf q} +h.c.)+\sum_{\mathbf q}\omega_{\mathbf q}(b^{\dag}_{\mathbf q}b_{\mathbf q}+1/2),
\label{Ham}
\end{eqnarray}
where $c_{\mathbf k}$ and $b_{\mathbf q}$ are the electron and phonon annihilation operators with momentum $\mathbf k$ and $\mathbf q$ respectively, $\epsilon(\mathbf k)$
and $\omega_{\mathbf q}$ are the electron band energy and the phonon frequency, and $N$ is the number of sites in the lattice. We also assume here that $\hbar=1$. In the following,
we consider dispersionless phonons with $\omega_{\mathbf q}=\omega_0$. The dimensionless matrix element of the electron-phonon interaction $\gamma(\mathbf q)$ has
different $\mathbf q$ dependence for different types of crystals. In ionic crystals (the Fr\"ohlich model \cite{Frelich}) with strong dispersion of the dielectric
permittivity in the long wavelength limit $|\gamma(\mathbf q)|^2=4\pi e^2/2\varkappa \omega_0 a^3q^2$, where $e$ is the elementary charge,
$\varkappa^{-1}=\varepsilon^{-1}_{\infty}-\varepsilon^{-1}_0$, $\varepsilon_0,\varepsilon_{\infty}$ are the static and the high frequency dielectric constants,
and $a$ is the lattice constant. In the case of molecular crystals the Holstein model is valid and $\gamma(\mathbf q)=g$ is momentum independent.

For the vast majority of crystalline solids, the adiabatic approximation is valid\cite{Ziman}. It means that the ratio of the electron band mass to the ionic mass is a
small parameter. Indeed, except for compounds with heavy fermions, the effective mass of electrons or holes is of the order of the free electron mass. Ion masses are at least
1700 times larger.  In organic solids, the applicability of the adiabatic approximation may be questionable. Indeed, organic materials which contain large molecules may have
relatively small hopping integrals $t\sim$ 0.1 eV because the molecular orbitals are spread over the whole molecule. On the other hand, intramolecular vibrations may be as
hard as 0.1-0.2 eV because of the presence of light atoms in the molecule.

Let us estimate the matrix elements in Hamiltonian Eq.(\ref{Ham}) in the Fr\"ohlich model \cite{Frelich}. The interaction between ions is dominated by the Coulomb attraction
$V(r)\sim e^2/\varepsilon_{\infty}r$. The phonon frequency is determined by the second derivative of $V(r)$ at $r=a$, $\omega_0\sim e/\varepsilon^{1/2}_{\infty}a^{3/2}M^{1/2}$.
The value of $\gamma(q)\omega_0 \sim \frac{t}{\sqrt{\varkappa}qa}\bigl (\frac{e^2ma}{\varepsilon_{\infty}} \bigr )^{1/2}\bigl (\frac{m}{M}\bigr )^{1/4}$. Here we use that
$t\approx 1/ma^2$. It is easy to see that the ratio $\frac{e^2ma}{\varepsilon_{\infty}}\approx$1. It means that everywhere in the Brillouin zone except the close vicinity
of $q=0$ point we have  the following hierarchy  $t:\gamma(\mathbf q)\omega_0:\omega_0\sim 1:(\frac{m}{M}\bigr )^{1/4}:(\frac{m}{M}\bigr )^{1/2}$. This hierarchy is also
valid in the case of the Holstein model. The electron-phonon matrix element in the Hamiltonian is always smaller than the kinetic energy of electrons. Nevertheless, from the
second and the third terms in Hamiltonian (\ref{Ham}) we can construct adiabatic (i.e. ion mass independent) energy
\begin{equation}
E_p=\frac{1}{N}\sum_{\mathbf{q}}|\gamma(\mathbf{q})|^2\omega_0,
\label{shift}
\end{equation}
which is called the polaron binding energy or the polaron shift. This energy does not depend on ionic mass and may be as large as $t$ or even larger. If following the works
of Eliashberg \cite{Eliashberg} we define the Eliashberg function and determine the dimensionless electron-phonon coupling constant it is determined exactly by
Eq. (\ref{shift}) $\lambda=E_p/zt$ ($z$ is the number of nearest neighbors in the lattice).

One can apply perturbation theory to Hamiltonian (\ref{Ham}). In the continuum approximation of the Fr\"ohlich model when $t\to \infty$ $a\to 0$ keeping the effective
mass $1/ta^2$ constant the perturbation theory is the expansion in the dimensionless parameter $\alpha={e^2\over{\varkappa}}\sqrt{m\over{2\omega_0}}$ \cite{Frelich,Smondyrev}.
In the lattice Holstein model the self energy represents the expansion in the dimensionless parameter $g^2\omega_0^2/t^2 \propto \sqrt{m/M}$ \cite{Barisic}. Some diagrammatic
expansions for the polaron self energy were used in the past in order to identify the instability associated with the polaron formation
\cite{AlexandrovKabanovRay,AlexandrovKabanov1996}. Nevertheless, within the standard perturbation theory the clear instability was not found.

\subsection{Translation symmetry broken ground state equations in the adiabatic limit $M\to\infty$}

The adiabatic theory was initially formulated for the Holstein model Eq.(\ref{Ham}) with $\gamma(\mathbf{k})=g$ in Ref.\cite{Holstein}. Later the theory was reformulated using
the field theory in the vicinity of nonzero classical solutions \cite{Rajaraman} in Ref.\cite{Kabanov}. This theory allows also us to analyze the nonadiabatic corrections to the
adiabatic solutions. In Ref.\cite{Kabanov} the equations which describe the saddle points of the adiabatic potential were derived. The
central equation in the adiabatic theory is the Schr\"odinger
equation for an electron moving in the external potential of the lattice
deformation. In the discrete lattice case (the tight binding approximation) it has the form \cite{Kabanov}:
\begin{equation}
-\sum_{{\mathbf m}\neq 0}t\psi_{\mathbf
{n+m}}^{l}+\sqrt{2} g \omega_{0} \varphi_{\mathbf n}
 \psi_{\mathbf n}^{l} = E_{l}\psi_{\mathbf n}^{l}.
 \label{shredinger}
\end{equation}
Here $\psi_{\mathbf n}$ is the electronic wave function on the site ${\mathbf n}$, $\varphi_{\mathbf n}$
is the deformation at the site ${\bf n}$, $l$ describes the quantum numbers of the problem, and the summation over ${\mathbf m}$ is taken over the nearest neighbours.
The important assumption of the adiabatic approximation is that the deformation field is very slow and we assume that $\varphi_{\mathbf n}$ is time independent,
$\partial\varphi/\partial t =0$  in Eq.(\ref{shredinger}). Therefore,
$\omega_{0}\propto M^{-1/2}\to 0$ and $g^{2}\propto M^{1/2}\to\infty$ but the polaron shift $g^{2}\omega_{0}=E_{p}$ is finite. The equation for $\varphi_{\mathbf n}$ has
the form \cite{Kabanov}:
\begin{equation}
\varphi_{\mathbf n}=-\sqrt{2} g|\psi_{\mathbf n}^{l}|^{2}.
\label{selfconsist}
\end{equation}
After substitution of Eq.(\ref{selfconsist}) to Eq.(\ref{shredinger}) we obtain:
\begin{equation}
-\sum_{{\mathbf m}\neq 0}t\psi_{\mathbf
{n+m}}^{l}-2 E_{p} |\psi_{\mathbf n}^{l}|^{2}
 \psi_{\mathbf n}^{l} = E_{l}\psi_{\mathbf n}^{l}.
 \label{nlinSch}
\end{equation}
Similar equations may be formulated in the case of Fr\"ohlich model \cite{Frelich}. In the continuum case, it was done by Pekar \cite{Pekar} and later for the discrete
lattice (see for example Refs.\cite{Kusmartsev,AlexandrovKabanov}). The first equation represents the Schr\"odinger equation for an electron in the potential $\phi_{\mathbf n}$
generated by the displaced ions:
\begin{equation}
-\sum_{{\mathbf m}\neq 0}t\psi_{\mathbf
{n+m}}^{l}-\sqrt{2}e \phi_{\mathbf n}
 \psi_{\mathbf n}^{l} = E_{l}\psi_{\mathbf n}^{l}.
 \label{shredingerfi}
\end{equation}
The equation for the potential $\phi_{\bf n}$ reads:
\begin{equation}
\varkappa\sum_{{\mathbf m}\neq 0}[\phi_{\mathbf n}-\phi_{\mathbf {n+m}}]=4\sqrt{2}\pi e |\psi_{\mathbf n}^{l}|^{2}.
\label{potential}
\end{equation}
Note that the left hand side of Eq.(\ref{potential}) represents the discrete version of
the Laplacian $\triangle$. If we solve Eq.(\ref{potential}) and substitute the solution back in to the Schr\"odinger equation we obtain exactly the equation derived by
Pekar \cite{Pekar}. Note, that Eqs.(\ref{shredinger},\ref{shredingerfi}) define the energy spectrum of the electron in the deformation $\varphi_{\mathbf n}$ or polarization
$\phi_{\mathbf n}$ fields. The total energy must include the positive energy of the deformation and the polarization itself. In the case of the Holstein model the deformation
energy is defined as $E_{def}=\omega_0\sum_{\mathbf n}|\varphi_{\mathbf n}|^2/2$.

Equation (\ref{nlinSch}) has two types of solutions. The first solution breaks the translation invariance and corresponds to the self-trapped state. This is the solution
that has nonzero deformation when the electron occupies the ground state level. The properties of the self-trapped solutions depend strongly on the system dimensionality.
In the 1D case, the self-trapped solution exists at any value of the polaron shift $E_p$. Therefore, the polaron is always stable in 1D. In 2D and 3D,  the self-trapped solution
exists only when the electron-phonon coupling constant $\lambda >\lambda_c$, where $\lambda_c$  is of the order of 1 and depends on the system dimensionality. The self-trapping
may occur with or without barrier formation depending on the value of the coupling constant $\lambda$ and the system dimensionality \cite{Emin,EminHolstein,Kabanov}.

The first correction to the adiabatic solution describes the renormalization of the local phonon mode. The electron self-trapping causes an increase of the local density near
the polaron center and leads to the shift of the local vibrational frequency. In the strong coupling limit $E_p\gg t$ the renormalized mode has the frequency \cite{Kabanov}:
\begin{equation}
\omega=\omega_0\sqrt{1-zt^2/2E^2_p},
\end{equation}
here $z$ is the number of nearest neighbours.

The second nonadiabatic correction describes the tunneling of self-trapped polarons.  The tunneling splitting was first derived by Holstein in Ref.\cite{Holstein}. A more
comprehensive formula was derived in Ref.\cite{AlexandrovKabanovRay} (see also Ref.\cite{Pasha4}). In the adiabatic limit the polaron tunneling is exponentially suppressed,
$t_{eff} \propto\sqrt{E_p\omega_0}\exp{(-g^2)} \propto E_p(m/M)^{1/4}\exp{(-const\sqrt{M/m})}$ (see Eq. (9) of Ref.\cite{AlexandrovKabanovRay}).

The second type of solution preserves translation invariance and represents itinerant band states. The deformation around the electron is absent $\varphi_{\mathbf n}=0$ therefore
all nonadiabatic corrections to this solution may be described in terms of ordinary perturbation theory and are described by the standard diagram technique, taking into account
that $m/M$ is small.

\subsection{The spectral function.}
As it follows from the previous section the eigenvalues and eigenstates of Hamiltonian (\ref{Ham}) in the adiabatic limit $m/M\to 0$ are determined by the two sets of states.
The first one is determined by a nonzero lattice deformation in the ground state and represents all eigenstates of the Hamiltonian in the presence of this deformation.
This type of eigenstate corresponds to a isystem with one polaron in the ground state of the grand-canonical Hamiltonian. The second set of eigenstates is the eigenstates of
the Hamiltonian without any lattice deformation. These states describe the empty system and electrons may appear in the system only due to an external perturbation like
photoexcitation or due to an injection. The parameter which controls the density of carriers is the chemical potential $\mu$. The spectral function is defined as
\cite{BonchBruevich,AGD}:
\begin{eqnarray}
A({\mathbf k},\omega)&=&{1\over{2\pi}}\Im\bigl (G_A({\mathbf k},\omega)-G_R({\mathbf k},\omega)\bigr )=
\nonumber \\& &-{1\over{\pi}}\Im G_R({\mathbf k},\omega),
\label{SpectrF}
\end{eqnarray}
here $G_R({\mathbf k},\omega)$, $G_A({\mathbf k},\omega)$ is the retarded and advanced Green's functions of an electron.  Using the Lehmann representation the spectral
function may be written as\cite{AGD,Mahan,Fetter}:
\begin{eqnarray}
A({\mathbf k},\omega)&=&{2\pi\over{\cal Z}}\sum_{n,m}|\langle n|c^{\dag}_{\mathbf k}|m\rangle|^2\delta(\omega+{\cal E}_m-{\cal E}_n)\nonumber\\
& &\bigl (\exp{(-\beta {\cal E}_n)}+\exp{(-\beta {\cal E}_m)}\bigr ),
\label{SpectrFT}
\end{eqnarray}
where $|i\rangle$ represents the eigenstates of the grand-canonical Hamiltonian $H^{'}=H-\mu N$ , with the eigenvalues ${\cal E}_{i}$, $H$ is defined in Eq.(\ref{Ham}),
$N=\sum_{\mathbf k}c^{\dag}_{\mathbf k}c_{\mathbf k}$ is the particle number operator and $\mu$ is the chemical potential. ${\cal Z}=\sum_n\exp{(-\beta {\cal E}_n)}$ is the
grand-canonical statistical sum, $\beta =1/k_BT$ ($k_B$ is the Boltzmann constant and $T$ is temperature). The zero temperature limit of this formula has a direct physical meaning:
\begin{eqnarray}
A({\mathbf k},\omega)&=&2\pi\sum_{n}\bigl [|\langle n|c^{\dag}_{\mathbf k}|0\rangle|^2\delta(\omega+{\cal E}_0-{\cal E}_n)\nonumber\\
& &+|\langle n|c_{\mathbf k}|0\rangle|^2\delta(\omega+{\cal E}_n-{\cal E}_0)\bigr ],
\label{SpectrF0}
\end{eqnarray}
here $|0\rangle $ and ${\cal E}_0$ are the ground state and the ground state energy of the grand-canonical Hamiltonian $H^{'}$. The first term in Eq.(\ref{SpectrF0}) describes
the inverse photo-emission spectrum, when the number of electrons in the ground state is increased by 1. Contrary, the second term describes the direct photo-emission spectrum,
when the number of electrons is reduced by 1.

Very often only the first term in Eq.(\ref{SpectrF0}) is calculated \cite{AlexandrovKabanovRay,Bonca1,Berciu,Bonca2} assuming that the chemical potential is large and
negative ($\mu<0$). Indeed, in that case, $|0\rangle$ is the phonon vacuum without any electrons, and therefore the second term in Eq.(\ref{SpectrF0}) is equal to zero.
Calculating eigenstates and eigenvalues of Hamiltonian Eq.(\ref{Ham}) with 1 electron it is easy to evaluate Eq.(\ref{SpectrF0}). On the other hand, the calculation of the
spectral function at arbitrary $\mu$ requires diagonalization of many-body grand-canonical Hamiltonian $H^{'}$ with an arbitrary number of electrons, which is quite difficult
task even for the exact diagonalization calculations \cite{AlexandrovKabanovRay,Bonca1,Bonca2}.

Note that the spectral density, calculated with the large negative chemical potential is not sensitive to the polaron states, which are described by the solutions of
Eq.(\ref{nlinSch}) with a nonzero lattice deformation in the ground state of the grand-canonical Hamiltonian. Indeed, the overlap of the wave functions with and without
the polaron deformation is exponentially small $\propto \exp{(-const\sqrt{M/m})}$. That is the reason why the sum rules for the spectral density do not contain any adiabatic
and nonadiabatic contributions, related to the polaron formation. To find polaronic features in this spectral function it is necessary to perform calculations with
exponential accuracy in the adiabatic parameter $m/M$.

In the next section, we present the calculations of the spectral function in the limit of large and negative chemical potential. Then we assume that chemical potential
is fixed in close vicinity to the polaron level in the way that the ground state has exactly one polaron. Here we consider spinless electrons and assume, that bipolaron
formation is suppressed.

\section{Results.}
\subsection{The spectral function for $\mu\to -\infty$ and the sum rules.}
In this section, we construct the spectral function for the Holstein model\cite{Holstein} in 1D, where the polaron represents the excited state of the grand canonical Hamiltonian
and the spectral function is given by the first term in Eq.(\ref{SpectrF0}). Here we neglect all exponentially small terms, i.e. the overlap between the wave functions with
zero and nonzero lattice deformation is neglected. Therefore, only terms with $\varphi_{\mathbf n}=0$ give nonzero contributions to the spectrum, given by Eq.(\ref{shredinger}).
This theory corresponds to the so-called sudden approximation. Note that it is exactly the reason why the sum rules do not have any contribution associated with the polaron
formation as mentioned in Ref.\cite{Berciu}. Solution of Eq.(\ref{shredinger}) with $\varphi_{\mathbf n}=0$ in 1D case gives the spectrum $\epsilon(k)=-2t\cos{(ka)}$, where $k$
is the electron momentum. Therefore, the spectral function in the adiabatic limit is given by the formula:
\begin{equation}
A(k,\omega)=\delta(\omega-\epsilon(k)).
\label{Sp_abs_adiabatic}
\end{equation}
This spectral function satisfies the sum rules $M_n=\int_{-\infty}^{\infty} \omega^nA(k,\omega)d\omega$ for $n=0,1$, derived by Kornilovich for the case of zero polaron
density $n \to 0$ \cite{Pasha3}. Here we do not write the chemical potential explicitly and absorb it to the definition of $\omega$. Note that all higher moments of the
spectral function contain explicitly the terms proportional to the electron-phonon coupling constant and are proportional to some powers of $(m/M)$. To demonstrate that, we
write the third sum rule $M_3=\epsilon(k)^3+2\epsilon(k)g^2\omega_0^2+g^2\omega_0^3$. Taking into account that $E_p=g^2\omega_0$ is independent of the ion mass $M$ we conclude
that the second term in the expression for $M_3$ is proportional to $(m/M)^{1/2}$ and the third term $\propto (m/M)$. Higher order moments $M_n$ have the mass independent
term $\epsilon(k)^n$ and the sum of $\sqrt{m/M}$ terms of powers $l$ where $1\leq l\leq n-1$. Therefore, if we neglect all nonadiabatic terms Eq.(\ref{Sp_abs_adiabatic})
satisfies all sum rules.

We can demonstrate that the spectral function of the Holstein model is indeed represented by the single $\delta$-function in the adiabatic limit by plotting in Fig. 1 the
spectral function of the two-site Hamiltonian $A(k=0,\omega)$ (See for example Ref.\cite{RanningerThibblin}).
This figure demonstrates that when $\omega_0/t\to 0$ the spectral density at $k=0$ is represented by a single peak centered at $-t$. At  $k=\pi/a$ the spectral function
is converging to a single peak at $\omega=t$. The width of this peak is decreasing to 0 when $\omega_0/t \to 0$. Since $M_0=1$, this corresponds to the definition of
the $\delta$-function.

\begin{figure}[tb]
\includegraphics[angle=0,width=1.0\linewidth]{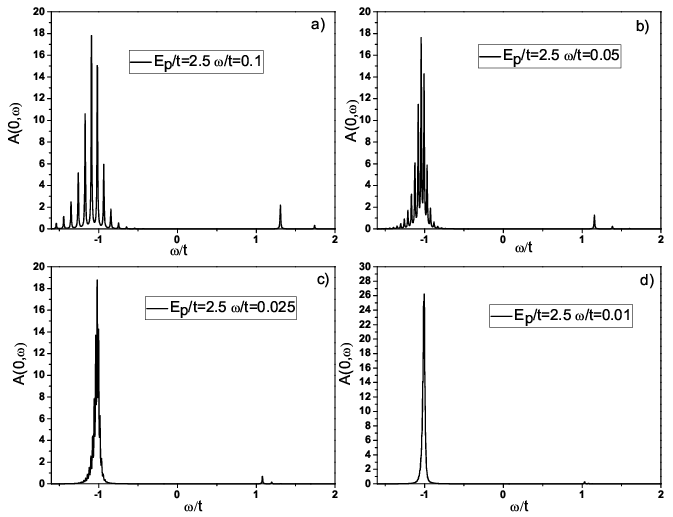}
\caption{The spectral function of the two-site Holstein model at $k=0$. When $\omega_0/t\to 0$ the spectral function converges to a single $\delta$-function.}
\label{fig:SP_abs}
\end{figure}

In the case when the ground state does not have polarons the solution of Eq.(\ref{selfconsist}) is trivial $\varphi_{\mathbf n}=0$.  The nonadiabatic correction to
the spectral function may be calculated by the standard perturbation theory. The dimensionless parameter, which determines the perturbation series is
$g\omega_0/2t=\sqrt{\lambda\omega_0/2t}\propto (m/M)^{1/4}$. Therefore, the perturbation theory may be considered as an expansion in series over the adiabatic
parameter $(m/M)^{1/2}$ because the expansion contains only even powers of the coupling constant. In order to calculate the spectral function with the given accuracy
$\delta$ it is necessary to calculate all irreducible diagrams for the self-energy up to the order $n$ which satisfies the inequality
$(g\omega_0/2t)^n=(\lambda\omega_0/2t)^{n/2} <\delta$. If we consider realistic parameters for crystalline materials $\omega_0/2t \le 0.1$ for required accuracy
better than $\delta=10$\% for $\lambda \le 3$ we have to take into account fourth order diagrams for the self-energy. Indeed, $\lambda^2(\omega_0/2t)^2\le0.09 < 0.1$.

Nevertheless, in the 1D Holstein model, the self-energy diverges near the threshold of the phonon emission. This divergence becomes even more pronounced in higher orders of
the perturbation theory. It is easy to see by the comparison of the contribution of the second-order and the fourth-order diagrams for the self-energy. The second order
contribution diverges near the threshold as $|x|^{-1/2}$ while the fourth order contribution diverges as $|x|^{-3/2}$, here $x=1-(\epsilon-\omega_0)/2t$ describes the deviation
from the threshold. It demonstrates that in order to describe the behaviour of the self-energy correctly the summation of all diagrams which contain the same divergencies
should be performed \cite{LevinsonRashba,Pitaevskii}.

\begin{figure}[tb]
\includegraphics[angle=0,width=0.3\linewidth]{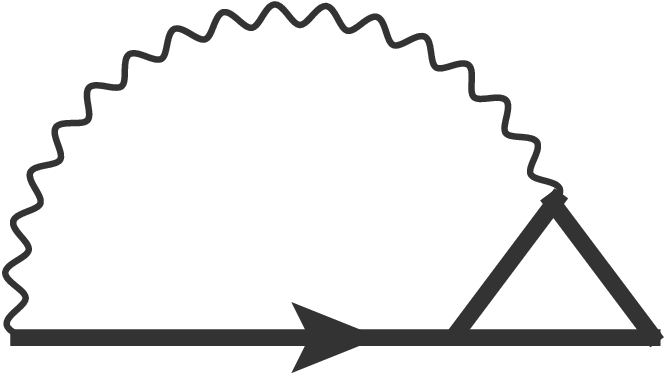}
\caption{Diagrammatic representation of Dyson equation with renormalized vertex.}
\label{fig:diagrams}
\end{figure}

\begin{figure}[tb]
\includegraphics[angle=0,width=1.0\linewidth]{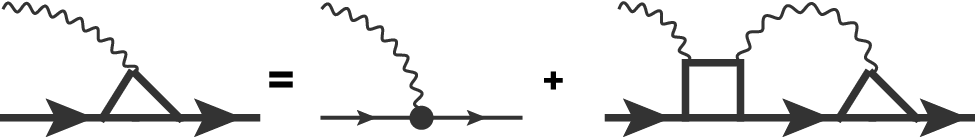}
\caption{Diagrammatic representation of the integral equation for the vertex. The kernel of the equation is defined by the square with two external electronic
lines and with two external phonon lines.}
\label{fig:vertex}
\end{figure}

\begin{figure}[tb]
\includegraphics[angle=0,width=1.0\linewidth]{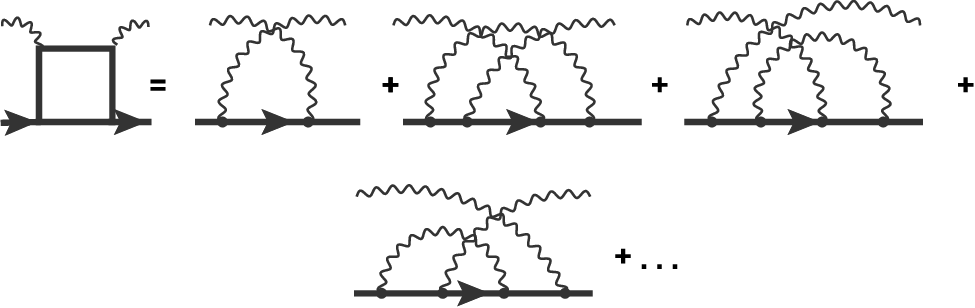}
\caption{Diagrammatic representation of the perturbation expansion for the square. This expansion is exact up to the sixth order in perturbation theory and
correctly accounting for the threshold effects.}
\label{fig:perturb}
\end{figure}

Following the procedure \cite{LevinsonRashba,Pitaevskii} we formulate Dyson's equation for the electron self-energy $\Sigma(\epsilon,k)$ (Fig.(\ref{fig:diagrams}) where
the vertex part $\Gamma(\epsilon,k,q)$ satisfies the equation schematically represented in Fig.(\ref{fig:vertex}). The kernel $K(\epsilon,k,q,x)$ of the integral equation
for the vertex part $\Gamma(\epsilon,k,q)$ is represented by the square, which contains all irreducible diagrams with one incoming and one outcoming electron lines and one
incoming and one outcoming phonon lines. The kernel may be evaluated by perturbation theory, schematically represented in Fig.(\ref{fig:perturb}). If we sum all diagrams
contributing to this kernel we present the exact solution of the problem with $\varphi_{\mathbf n}=0$. The exponentially small terms corresponding to the overlap of the states
with and without deformation cannot be evaluated in this procedure because of the non-analytic nature of these terms. The diagrams shown in Fig.(\ref{fig:perturb}) allow
evaluating vertex (Fig.(\ref{fig:vertex}) which is exact in sixth order of perturbation expansion and correctly describes the threshold effects. Finally, the solution
of Dyson's equation (Fig.(\ref{fig:diagrams})) with the vertex defined by Fig.(\ref{fig:vertex}) and with the kernel defined by diagrams, plotted in Fig.(\ref{fig:perturb})
represents the summation of all diagrams, where the number of crossings in each phonon line is less than four. This theory is accurate up to the sixth order in the dimensionless
parameter $(\lambda\omega_0/2t)^{1/2}\propto (m/M)^{1/4}$ and correctly describes the behaviour of the self-energy near the phonon emission threshold. Therefore, the accuracy
of our calculations for $\lambda<3$ and $\omega_0 \leq 0.2t$ is better than 3\%.

After integration over energies Dyson's equation represented in Fig.(\ref{fig:diagrams}) and the equation for the vertex part represented in Fig.(\ref{fig:vertex}) have the form:
\begin{equation}
\Sigma(\epsilon,k)={g\omega_0\over{2\pi}}\int_{-\pi}^{\pi}\Gamma(\epsilon,k,q)G(\epsilon-\omega_0,k-q)dq,
\label{eq:Dyson}
\end{equation}
and
\begin{equation}
\Gamma(\epsilon,k,q)=g\omega_0+\int_{-\pi}^{\pi}K(\epsilon,k,q,x)
G(\epsilon-\omega_0,k-x)\Gamma(\epsilon,k,x){dx\over{2\pi}}.
\label{eq:vershina}
\end{equation}
The kernel $K(\epsilon,k,q,x)$ calculated up to the sixth order in perturbation theory (Fig.(\ref{fig:perturb})) is represented by the equation:
\begin{eqnarray}
&&K(\epsilon,k,q,x)=(g\omega_0)^2G(\epsilon-2\omega_0,k-q-x)+
(g\omega_0)^4\int_{-\pi}^{\pi}G(\epsilon-3\omega_0,k-q-x-y)\times \Bigl ( \nonumber\\
&&G(\epsilon-2\omega_0,k-q-y)G(\epsilon-2\omega_0,k-x-y)+
G(\epsilon-2\omega_0,k-q-x)G(\epsilon-2\omega_0,k-x-y)+\nonumber\\
&&G(\epsilon-2\omega_0,k-q-y)G(\epsilon-2\omega_0,k-q-x)\Bigr ){dy\over{2\pi}}.
\label{eq:jadro}
\end{eqnarray}
Here the electron Green's function is defined as:
\begin{equation}
G(\epsilon,k)={1\over{\epsilon+2t \cos{(k)}-\Sigma(\epsilon,k)}}.
\label{eq:FG}
\end{equation}
As a result, the problem of calculation of the spectral function is reduced to a solution of two integral equations for the self-energy Eq.(\ref{eq:Dyson}) and for the vertex
part Eq.(\ref{eq:vershina}) with the kernel, defined by Eq.(\ref{eq:jadro}). Here we present the numerical solution to these equations. We also propose an accurate approximation
for the vertex part and compare the numerical solution with the approximate solution.

In order to solve Eqs.(\ref{eq:Dyson},\ref{eq:vershina}) the energy was defined in 3001 points on the interval $-6t < \epsilon < 6t$. The momentum was defined in 101 points
of the Brillouin zone $-\pi < k < \pi$ with the step $\Delta h=2\pi/100$. The integration was performed by the trapezoidal rule with the accuracy $\sim (\Delta h)^2$ better
than 1\%. As a staring point of the iteration procedure, we introduce the vertex and the self-energy which are averaged over momenta and represent the solution of
Eqs.(\ref{eq:Dyson},\ref{eq:vershina}) averaged over momenta
\begin{eqnarray}
\Gamma_{av}(\epsilon)=g\omega_0/\Bigl[1-{{g^2\omega_0^2\over{t^2}}(1+{{g^2\omega_0^2\over{t^2}}\over{[(z_2-z_2^{-1})(z_3-z_3^{-1})]}})\over{[(z_1-z_1^{-1})(z_2-z_2^{-1})]}}\Bigr].
\label{eq:gamma_av}
\end{eqnarray}
Here $z_n$ is the smallest ($|z_n|\leq 1$) root of the quadratic equation $z^2+\epsilon_n/t z+1=0$, and $\epsilon_n=\epsilon-n\omega_0-\Sigma(\epsilon-n\omega_0)$. The self-energy
$\Sigma(\epsilon)$ is defined from Eq.(\ref{eq:Dyson}) with approximate vertex $\Gamma_{av}(\epsilon)$. Note, that the Green's function Eq.(\ref{eq:FG}) with this self-energy
satisfies the sum rules up to the seventh order.

\begin{figure}[tb]
\includegraphics[angle=0,width=1.0\linewidth]{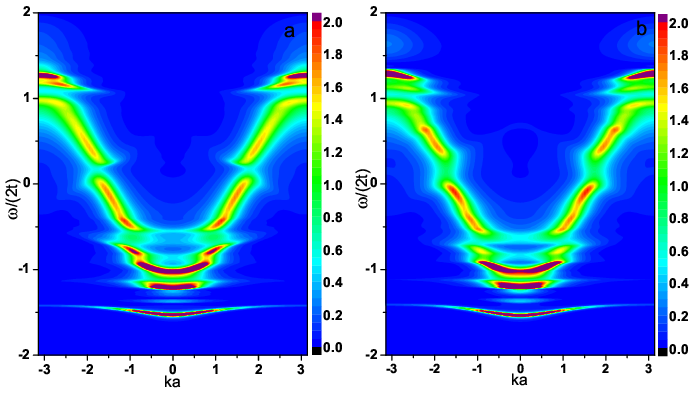}
\caption{2D plot of the spectral function of the 1D Holstein model in the limit $\mu\to -\infty$. $E_p/2t=2.5$ and $\omega/2t=0.1$. Colors represent the value of
the spectral function at given energy and momentum. Exact numerical solution of Eqs.(\ref{eq:Dyson},\ref{eq:vershina}) (a), and approximation Eq.(\ref{eq:gamma_app2})
for the vertex part (b).}
\label{fig:SP-fun-1D_1}
\end{figure}

The approximate self-energy and vertex part are used as a starting point to solve Eqs.(\ref{eq:Dyson},\ref{eq:vershina}) iteratively. The iteration procedure looks as follows.
On every step the self-energy is calculated from Eq.(\ref{eq:Dyson}) with the vertex from the previous step. Then the new self-energy is used to obtain the new vertex from
Eq. (\ref{eq:vershina}). Dyson's equation Eq. (\ref{eq:Dyson}) was solved by iterations.  On the other hand, the standard routine for the solution of the system of the linear
equation from the NAG library was used in every step of the iteration. Usually, only a few iterations are necessary to obtain solutions for the vertex part and the self-energy
Eqs.(\ref{eq:Dyson},\ref{eq:vershina}). The results of calculations are presented in Fig(\ref{fig:SP-fun-1D_1}a).

\begin{figure}[tb]
\includegraphics[angle=0,width=1.0\linewidth]{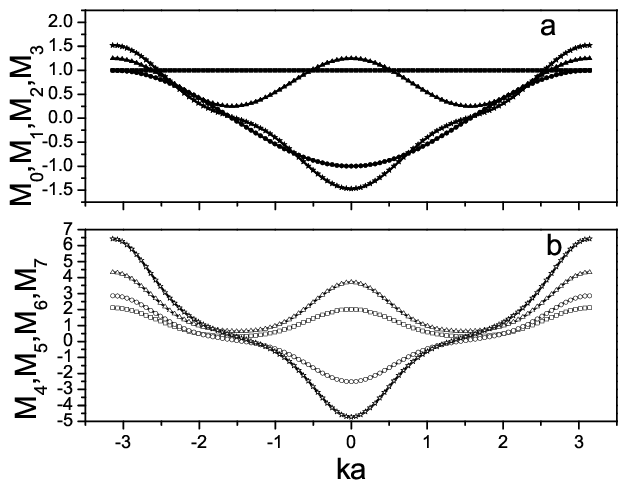}
\caption{Moments of the spectral function calculated numerically (symbols) in comparison with exact sum rules (lines) for a) $M_0$ full squares, $M_1$ full circles,
$M_2$ full triangles, $M_3$ full stars, b) $M_4$ empty squares, $M_5$ empty circles, $M_6$ empty triangles, and $M_7$ empty stars.}
\label{fig:SP-fun-1D-sum-rules}
\end{figure}

To construct an approximate solution for the vertex $\Gamma(\epsilon,k,q)$ we notice first, that the main contribution to the integral in Eq.(\ref{eq:vershina}) comes
from the vicinity of $x=k$ point. Note, that $k-x=0$ point represents the Van Hove singularity in Eq. (\ref{eq:vershina}), because at this point group velocity is equal to 0. Therefore, we may take out the vertex part $\Gamma(\epsilon,k,k)$ from the integral. As a result, we obtain:
\begin{equation}
\Gamma_{app}(\epsilon,k,q)=g\omega_0+\Gamma_{app}(\epsilon,k,k)F(\epsilon,k,q),
\label{eq:gamma_app1}
\end{equation}
where $F(\epsilon,k,q)=\int_{-\pi}^{\pi}K(\epsilon,k,q,x)G(\epsilon-\omega_0,k-x){dx\over{2\pi}}$
Rewriting this equation at $q=k$ and then solving equation for $\Gamma(\epsilon,k,k)$, we obtain approximation for the vertex:
\begin{equation}
\Gamma_{app}(\epsilon,k,q)=g\omega_0+g\omega_0F(\epsilon,k,q)/(1-F(\epsilon,k,k)).
\label{eq:gamma_app2}
\end{equation}
Integrals in formula for $F(\epsilon,k,q)$ may be calculated  analytically (See Appendix A). Substituting this vertex with analytic formulae for $F(\epsilon,k,q)$
Eqs.(\ref{eq:F2},\ref{eq:F31},\ref{eq:F32},\ref{eq:F33}) to Dyson equation Eq.(\ref{eq:Dyson}) we calculate spectral function, presented in Fig.(\ref{fig:SP-fun-1D_1}b).
Comparison of spectral functions presented in Fig.(\ref{fig:SP-fun-1D_1}a) and Fig.(\ref{fig:SP-fun-1D_1}b) demonstrates that indeed the main contribution to the vertex
$\Gamma(\epsilon,k,q)$ is coming from the vicinity of the Van Hove singularity in Eq.(\ref{eq:vershina}). Therefore, there is very good agreement between numerical solution
of Eqs.(\ref{eq:Dyson},\ref{eq:vershina},\ref{eq:jadro}) and approximate results for the spectral function. Note, that both spectral functions satisfy the sum rules up to
the seventh moment (Fig(\ref{fig:SP-fun-1D-sum-rules})).

Fig.(\ref{fig:SP-fun-1D_1}) clearly demonstrates that the lowest energy band has minimum at $\epsilon=\epsilon(k=0)\propto (E^2_p\omega^2_0/t)^{1/3} \sim (m/M)^{1/3}$. With
increasing $k$ the energy disperses relatively quick up to the energy of the order of $\epsilon(k)=\epsilon(k=0)+\omega_0$ at $k\approx \sqrt{2m\omega_0}$ and then remains
unchanged with further increase of $k$ up to the Brillouin zone edge. This effect is well known in literature \cite{LevinsonRashba} and called the threshold phenomenon. Note
that the threshold phenomena were first discussed in connection with the excitation spectrum of the superfluid $He$ by Pitaevskii \cite{Pitaevskii}. In the case of the 1D Holstein
model, the anomaly in the self-energy belongs to type "c" as discussed by Levinson and Rashba \cite{LevinsonRashba} and more recently by Goodvin and Beciu \cite{Goodvin2010}.
The self-energy shows a strong divergence $-(-2t+\omega_0-\epsilon)^{-1/2}$ when $\epsilon\to-2t+\omega_0-0$. This first branch of the spectrum does not have any damping,
because any inelastic scattering of electrons is forbidden by the conservation of energy.
The anomalies similar to the threshold effect are also present at higher energies because the higher order diagrams have additional divergence at $-2t+n\omega_0$,
$n=1,2,3,...$. These anomalies are much less pronounced because of the finite imaginary part of the self-energy at these energies. And finally, there is an increase of
$A(k,\omega)$ near $ka=\pi$ and $\omega>2t$. This peak is less pronounced than the peak at the bottom of the band, because it has finite damping.

%\begin{figure}[tb]
%\includegraphics[angle=0,width=1.0\linewidth]{Fig_diagrams.eps}
%\caption{Infinite series of diagrams which correspond to the integral equation for the electron-phonon vertex.}
%\label{fig:series}
%\end{figure}

\begin{figure}[tb]
\includegraphics[angle=0,width=1.0\linewidth]{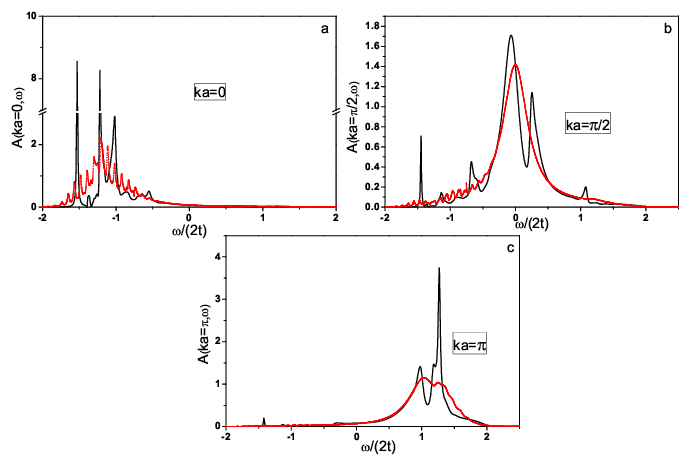}
\caption{The spectral function of the 1D Holstein model in the limit $\mu\to -\infty$ at different points of the Brillouin zone. Dotted lines represent the spectral function
calculated within the momentum averaged approximation with corrections of level two. $E_p/2t=2.5$ and $\omega/2t=0.1$.}
\label{fig:SP-fun-1D_2}
\end{figure}

Now let us compare the present theory, which takes into account all diagrams up to the sixth order and represents the partial summation of infinite perturbation series with the momentum
average approximation, proposed in Ref.(\cite{Berciu}). This theory is accurate in the second-order approximation. The fourth-order contribution in this theory is approximated
by the self-energy contribution:
\begin{equation}
\Sigma_4(k,\epsilon)\approx 2{g^4\omega_0^4\over{(2t)^3}}I^2(z_1)I(z_2),
\label{eq:Berciu}
\end{equation}
which is equal to an averaged over $k$ fourth order diagram with phonon lines crossings (\ref{eq:sig4C}), and $I(z)=2/(z-z^{-1})$.
This expression should be compared with the exact forth order contribution $\Sigma_4(k,\epsilon)=\Sigma_4^{NC}(k,\epsilon)+\Sigma_4^{C}(k,\epsilon)$, where
\begin{equation}
\Sigma_4^{NC}(k,\epsilon)=4{g^4\omega_0^4\over{(2t)^3}}{z_1+1/z_1\over{(z_1-1/z_1)^3}}I(z_2),
\label{eq:4NC}
\end{equation}
%\begin{widetext}
\begin{equation}
\Sigma_4^{C}(k,\epsilon)={g^4\omega_0^4\over{(2t)^3}}{2(1-z_1^4z_2^2)\over{z_1^4z_2^2+1-2z_1^2z_2cos(k)}}I(z_1)^2I(z_2),
\label{eq:sig4C}
\end{equation}
%\end{widetext}
here $z_1$ and $z_2$ ($|z_{1,2}|<1$) roots of quadratic equations, defined after Eq.(\ref{eq:gamma_av}) and are calculated with $\Sigma_B(\omega)=-i\delta$. Comparing these
results we conclude that momentum average approximation is rather poor because it provides the self-energy which is functionally different from the perturbation theory.
Therefore, momentum average approximation is accurate only in the second order in coupling constant. There were few attempts to improve this approximation \cite{BerciuGoodvin}.
The first step of corrections leads to the correct expression for fourth order noncrossing diagram $\Sigma_4^{NC}(k,\omega)$ Eq.(\ref{eq:4NC}) but fails to take into account
$k$-dependence of the diagram with crossing of phonon lines $\Sigma_4^{C}(k,\epsilon)$. The next and the last level of corrections leads to some infinite system of inhomogeneous
equations, which was solved by truncation. This procedure is not well justified because the coefficient in this equation does not fall with the increasing of the system size.
Nevertheless, the solution of this system expanding in powers of the coupling constant may be performed analytically and we recover the correct expression for the fourth-order
diagrams.  Higher order diagrams require the next level of corrections which were not discussed in Ref.\cite{BerciuGoodvin}. Therefore, we conclude that momentum average
approximation with the corrections of levels 1 and 2 is accurate up to the fourth order in perturbation theory.

In Fig.(\ref{fig:SP-fun-1D_2}) the spectral function is plotted for three different $ka=0,\pi/2,\pi$. In the same graphs, the spectral functions calculated within momentum
average approximations with the corrections of level 2 \cite{BerciuGoodvin} are presented. There is quite good agreement between spectral functions calculated within these
approaches at large momenta ($ka=\pi/2,\pi$).  At $ka=0$ the incoherent part of the spectrum is also quite similar in both cases. Nevertheless, the present theory shows much
sharper peaks. These spectral functions satisfy the sum rules up to 7th moment (Fig(\ref{fig:SP-fun-1D-sum-rules})).

\subsection{The spectral function in the polaron state}
In order to describe the spectral function of the system where the ground state of the grand canonical Hamiltonian $H^{'}$ corresponds to a finite polaron density, we have
to tune the chemical potential close to the single polaron level. The requirement is that the lowest energy state of the grand canonical Hamiltonian $H^{'}$ with at least
one polaron should be lower than the lowest energy level of the Hamiltonian without polarons. In order to prevent bipolaron formation, we consider the spinless fermions
and assume that bipolarons represent an excited state on the grand canonical Hamiltonian when the chemical potential is pinned near the polaron level.

 The spectral function Eq. (\ref{SpectrF0}) has two terms. The first term describes the creation of an additional carrier in the ground state of the grand canonical Hamiltonian
 $H^{'}$ with one polaron which is proportional to $1-n$, where $n=1/N$ is the polaron density. More important is the second term which describes the emission of the electron
 from the ground state which is proportional to $n=1/N$. Importantly, exactly this term is measured in direct photoemission spectroscopy. Note that the second term carries
 information about polaron formation. This was pointed out in Ref.\cite{Pasha3} where the exact formula for the first moment of the spectral function was derived. This first
 moment has the contribution, which is proportional to the adiabatic polaron shift $2 n E_p$. Therefore, this term may be derived from the adiabatic theory without involving
 nonadiabatic corrections.

In the adiabatic approximation, the polaron state is the self-trapped state localized in the translation symmetry broken deformation field $\varphi_{\mathbf n}$. This state
is degenerate because the polaron energy does not depend on the polaron position and is described by the solution of Eq.(\ref{nlinSch}) $\psi_{\mathbf n}^0$ with the lowest
energy $E_0$. Within the sudden approximation the second term in Eq.(\ref{SpectrF0}) is
proportional to the square of the Fourier transform of the  $F_k=\sum_n\exp{(ikn)}\psi_{\mathbf n}^0/\sqrt{N}$. Therefore, Eq.(\ref{SpectrF0}) is given by a single $\delta$
function:
\begin{eqnarray}
A_-({\mathbf k},\omega)&=&2\pi \sum_{l}|\langle l|c_{\mathbf k}|0\rangle|^2\delta(\omega+{\cal E}_l-{\cal E}_0)\nonumber\\
&=&2\pi |F_k|^2 \delta(\omega-E_0+\mu).
\label{Spectr_emis}
\end{eqnarray}
Since the polaron wave function is localized $F_k\propto 1/N^{1/2}$ the spectral function $A_-(k,\omega)$ is proportional to polaron density $N^{-1}$ (one polaron per $N$ sites).
The chemical potential $\mu$ is pinned to the polaron energy $\mu=E_0+E_p\sum_n|\psi_n^0|^4$ \cite{Kabanov} which is the sum of the lowest eigenvalue of Eq(\ref{nlinSch}) and
the deformation energy caused by the polaron. In the strong coupling limit $\mu=-E_p-t^2/E_p$ therefore the peak of the spectral function $A(k,\omega)$ is at energy
$\omega=-E_p+t^2/E_p$. Note, that the spectral function is equal to 0 at the chemical potential ($\omega=0$). This is because the overlap of the wave functions with and without
polaron deformation is exponentially small and the annihilation of the polaron  together with the deformation is forbidden due to the Franck-Condon principle. The part of the
spectral function associated with the inverse photo-emission is given by the first part of Eq.(\ref{SpectrF0}) and may be written as
\begin{eqnarray}
A_+({\mathbf k},\omega)&=&2\pi\sum_{n}|\langle n|c_{\mathbf k}^\dagger|0\rangle|^2\delta(\omega+{\cal E}_0-{\cal E}_n)\nonumber\\
&=&2\pi\sum_{l\ne 0}|F_k^l|^2 \delta(\omega-E_l+\mu),
\label{Spectr_absor}
\end{eqnarray}
where $F_k^l=\sum_n\exp{(ikn)}\psi_{\mathbf n}^l/\sqrt{N}$ is the Fourier transform of the wave function which represents the $l$-th eigenstate of Eq.(\ref{shredinger})
where $\varphi_{\mathbf n}$ is determined by Eq.(\ref{selfconsist}) with $\psi_{\mathbf n}^l=\psi_{\mathbf n}^0$. Here the sum does not include $l=0$ because the absorption
of electron directly to the polaronic state is proportional to the overlap of the lattice wave functions with zero and nonzero deformation which is exponentially small
$\propto \exp{(-\sqrt{M/m})}$. As in the photoemission case this process is forbidden due to the Franck-Condon principle. Note that the spectrum has a clear pseudogap,
which corresponds to $|E_0|-2t\approx 2E_p-2t$. The physics of this pseudogap is simple. The probability to remove polaron from the Fermi level or the probability to add polaron
to the Fermi level is exponentially suppressed and these processes are too weak to observe them experimentally. Therefore, the main features of the spectral function are
shifted from the Fermi energy because of the Franck-Condon principle.

\begin{figure}[tb]
\includegraphics[angle=0,width=1.0\linewidth]{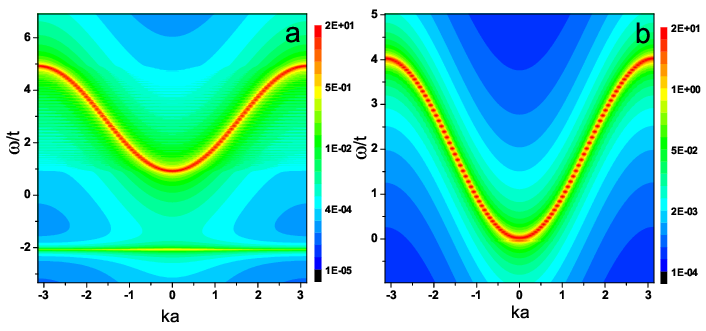}
\caption{2D plot of the polaron spectral function in the adiabatic limit for 1D chain with $N=100$ and periodic boundary conditions. Polaron density is therefore $n=0.01$.
Colors represent the value of the spectral function at given energy and momentum. a) $E_p=2.5t$, $E_0=-4.9808t$ and $\mu$ is pinned by polaron level at $-2.9087t$. b)
$E_p=0.5t$, $E_0=-2.0633t$ and the chemical potential $\mu=-2.0210t$.}
\label{fig:SP-polaron-1D}
\end{figure}

In Fig.(\ref{fig:SP-polaron-1D}) the spectral function of the polaron is plotted for the 1D system with 100 sites with periodic boundary conditions in the adiabatic limit for
the polaron binding energy $E_p=2.5t$ Fig.(\ref{fig:SP-polaron-1D}a) and for $E_p=0.5t$ Fig.(\ref{fig:SP-polaron-1D}b). The main spectral features repeat the free electron
spectrum except that the presence of polaron deformation breaks the translation invariance. Therefore, the linewidth has finite broadening due to the scattering of a free
electron on the polaron deformation. As expected, this spectral density satisfies the zero $M_0=1$ and the first $M_1=-2t cos(ka)-2nE_p$ sum rules, $a$ is the lattice
constant and $n=1/N$ is the polaron density.

In the limit when the polaron size is larger than the lattice constant Eqs.(\ref{shredinger},\ref{selfconsist},\ref{nlinSch}) have analytic solutions for both the localized
and itinerant part of the spectrum (See Appendix B). Using these solutions the matrix elements $F_k$ and $F_k^l$ in Eqs.(\ref{Spectr_emis},\ref{Spectr_absor}) may be evaluated
analytically:
\begin{equation}
F_k=\pi\sqrt{t\over{E_p N}}{1\over{\cosh{(\pi kat/E_p)}}},
\label{Fk}
\end{equation}

\begin{eqnarray}
F_{k}^l&=&i(({2lat\over{E_p}})^2+1-{4t\over{NE_p}})^{-1/2}\nonumber\\
& &\Bigl [-({4lat\over{E_p}}){\sin{(\kappa aN/2)}\over{\kappa aN}}+{4\sin^2{(\kappa aN/4)}\over{\kappa aN}} +\nonumber \\
&-&{2\over{\kappa aN}}+{2\pi t\over{E_p N}}{1\over{sinh{(\pi\kappa at/E_p)}}}\Bigr ].
\label{Fkk}
\end{eqnarray}
Here $a$ is the lattice constant, $N$ is the number of sites in the system, $\kappa=l-k$, where $l$ represents the solutions of Eq.(\ref{per_condit}) $k=2\pi n/Na$ where
$n=0,1,2,.,N-1$. These two expressions substituted to Eqs.(\ref{Spectr_emis},\ref{Spectr_absor}) with $E_0=-2t-E_p^2/4t$, $E_l=-2t+tl^2a^2$, and the chemical potential
$\mu=-2t-E_p^2/12t$ provides the analytic description of the spectral function which is accurate up to the nonadiabatic corrections which are small as $\sqrt{m/M}$. Note
that this analytic expression satisfies the exact sum rules (i.e. $M_0$ and $M_1$)\cite{Pasha3} when the polaron density is finite.

\begin{figure}[tb]
\includegraphics[angle=0,width=1.0\linewidth]{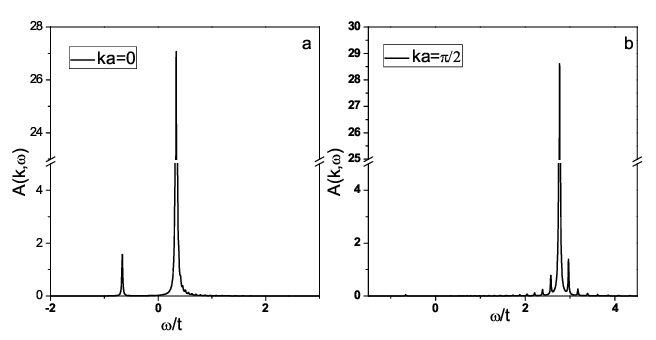}
\caption{The polaron spectral function in the adiabatic limit for the 1D chain with $N=100$ in the continuous limit and the polaron density $n=0.01$. $E_p=2t$ and the
chemical potential $\mu=-2t-t/3$ and the polaron binding energy $E_0=-3t$  a) $ka=0$. b) $ka=\pi/2$.}
\label{fig:SP-polaron-D-analitic}
\end{figure}

The spectral function $A(k,\omega)$ calculated on the basis of Eqs.(\ref{Spectr_absor},\ref{Spectr_emis}) using Eqs.(\ref{Fk},\ref{Fkk}) for two different momenta is plotted
in Fig.\ref{fig:SP-polaron-D-analitic}.
As expected the results are very similar to that plotted in Fig.\ref{fig:SP-polaron-1D}. There is a weak spectral intensity at $E_0-\mu$. This intensity is proportional to
the polaron density and is suppressed quickly when momentum moves away from the $k=0$ point. Main intensity is centered at $\omega=E_p^2/12t+tk^2a^2$. Because the exact wave
functions Eq.(\ref{WF-excited}) are not plane waves there is a natural broadening of the spectral line due to the scattering of an electron on the polaron deformation. Therefore,
the line, corresponding to an almost free electron has finite width in energy. This broadening is very important because it compensates the contribution from the polaron state
at negative energies in the sum rules. Therefore, the exact sum rules $M_0=1$ and $M_1=-2nE_p+tk^2a^2+E_p^2/12t$ \cite{Pasha3} are satisfied.

Note that in Ref. (\cite{Pasha3}) the sum rule $M_2$ for the spectral function was derived (See Eq.(15) in \cite{Pasha3}). This moment contains averages of
$\langle\varphi_n\rangle$ and $\langle\varphi_n^2\rangle$, which cannot be evaluated exactly. Nevertheless, within the adiabatic approximation, these averages may be easily
evaluated. Indeed, Eq.(\ref{selfconsist}), defines the deformation field at site $\mathbf{n}$. Therefore, integration of this equation with $\psi^0(x)$ defined by
Eq.(\ref{gr_State}) leads to the result, obtained in \cite{Pasha3} $\langle\varphi\rangle=-\sqrt{2}g$ which corresponds to the case of one polaron per $N$ sites. This
immediately leads to the correct equation for the momentum $M_1$. Similarly, $\langle\varphi^2\rangle=2g^2\sum_n|\psi_n^0|^4$. Calculation of the sum leads to the following
result $\langle\varphi^2\rangle=g^2E_p/3t$ in the weak coupling case $E_p<<t$ and $\langle\varphi^2\rangle=g^2(1-t^2/E_p^2)$.
Therefore, the second moment of the spectral function has the form:
\begin{widetext}
\begin{equation}
M_2=(\epsilon(k)-\mu)^2-4nE_p(\epsilon(k)-\mu)+\begin{cases}
2nE_p^3/3t & \textrm{if $E_p<<t$}\\
4n(E_p^2-t^2) & \textrm{if $E_p>>t$}
\end{cases}
+{\cal O}(\sqrt{m/M}).
\label{M2}
\end{equation}
\end{widetext}

\begin{figure}[b]
\includegraphics[angle=0,width=1.0\linewidth]{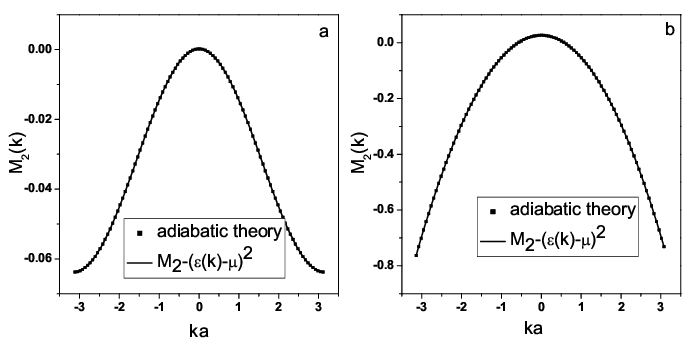}
\caption{The second moment of the spectral function as a function of the momentum $k$ calculated in the adiabatic limit.    a) Tight binding calculations with $E_p=0.4t$
and $\mu= -2.0134t$. b) Analytic solution of Eq.(\ref{Sch_deform}) with $E_p=2t$ and $\mu=-2.3333t$.}
\label{fig:sum_rule}
\end{figure}

In Fig.\ref{fig:sum_rule} the calculated sum rule $M_2$ is plotted as a function of momentum $k$ in comparison with expression Eq.(\ref{M2}). The calculations are performed
for the tight binding model (Eq.(\ref{shredinger},\ref{selfconsist})) and for analytic differential equation (\ref{Sch_deform}). In both cases the results perfectly match
the analytic formula Eq.(\ref{M2}).

\section{Conclusion}
We formulate the analytic theory for the spectral function of the electron-phonon system in the adiabatic limit. In the limit of the dilute polaron system where the
polaron density $n\to 0$ and when the chemical potential is large and negative $\mu \to -\infty$ the polaron deformation is absent and the spectrum of electrons
coupled to phonons is defined by the standard perturbation theory. Using the diagram technique with an accurate account of threshold effects we were able to formulate
an integral equation for the vertex, which takes into account all diagrams up to the sixth order. Then we solve Dyson's equation for the self-energy with the renormalized
vertex and calculate the polaron spectral function in 1D with an accuracy better than 3\% for $\lambda \leq 3$ in the adiabatic limit. The moments of the spectral function
satisfy the exact sum rules up to the 7th moment.

For the system with a finite polaron density when the chemical potential is pinned at the polaron level the ground state has nonzero adiabatic deformation due to
the presence of polarons. In this case, the spectral function is calculated in the adiabatic limit without any non-adiabatic corrections. We also show how the adiabatic
terms, proportional to the polaron density $n$, may be evaluated for higher-order moments. The spectral function shows weak spectral intensity at $E_0-\mu$, which is
proportional to $n$. At the Fermi level in the adiabatic limit the spectral density is absent. Continuous spectrum starts at $-\mu-2t>0$. Therefore, the spectrum has a gap,
which is proportional to the polaron binding energy $E_0$. Contrary to the case of $n=0$ the spectral function with a finite polaron density has some adiabatic contributions
which are characteristic of the polaron formation.

\section{ACKNOWLEDGMENTS}
The author thanks D. Mihailovic, T. Mertelj and especially P.E. Kornilovitch for very helpful and enlightening discussions and acknowledges the financial
support from the Slovenian Research Agency Program No. P1-0040.

\appendix

\section{Appendix A}
\subsection{Derivation of analytical formula for approximate vertex.}
The integrals in the expression for function $F(\epsilon,k,q)$ may be evaluated analytically. After the substitution $z=\exp{(ik)}$ the integral becomes
$\int_{-\pi}^{\pi}dk/2\pi \to \oint_{|z|=1} dz/2i\pi z$ and is determined by the poles of the integrand within the unite circle $|z|=1$.
The residues are determined by the quadratic equation: $z^2-\epsilon_nz/t+1=0$, which has two roots $z_n$ ($|z_n|\leq 1$) and $1/z_n$.
Therefore, all integrals for $F(\epsilon,k,q)$ are expressed in terms of $z_n$. Here $\epsilon_n=\epsilon-n\omega_0-\Sigma(\epsilon-n\omega_0)$ and
$\Sigma(\epsilon)$ is the solution of Dyson equation (\ref{eq:Dyson}) determined with $\Gamma_{av}(\epsilon)$ Eq.(\ref{eq:gamma_av}).
\begin{eqnarray}
F(\epsilon,k,q)&=&F_2(\epsilon,k,q)+F_{4,1}(\epsilon,k,q)+\nonumber \\
&&F_{4,2}(\epsilon,k,q)+F_{4,3}(\epsilon,k,q),
\label{eq:F}
\end{eqnarray}
where $F_2$ represents second order contribution, and $F_{4,i}$, $i=1,2,3$ represent three contributions of forth order in accordance with Eq.(\ref{eq:jadro})

The second-order contribution has the form:
\begin{widetext}
\begin{equation}
F_2(\epsilon,k,q)={4g^2\omega_0^2\over{(z_1-z_1^{-1})(z_2-z_2^{-1})}}\Big[{z_1z_2\over{\mu-z_1z_2}}-{z_1^{-1}z_2^{-1}\over{\mu-z_1^{-1}z_2^{-1}}}\Big].
\label{eq:F2}
\end{equation}
Three fourth order terms after cumbersome calculations have the following form:
\begin{eqnarray}
F_{3,1}(\epsilon,k,q)&=&{16g^4\omega_0^4\over{(z_1-z_1^{-1})(z_2-z_2^{-1})^2(z_3-z_3^{-1})}}\Big[{\eta z_2z_3\over{\mu-z_2z_3}}\big({z_1z_2^2\over{\mu-\eta z_1z_2^2}}
+{1\over{\eta-z_1z_2z_3}}\big)+{\eta z_1z_2z_3\over{(\eta-z_1z_2z_3)(\eta-\mu z_1z_2^2)}}\nonumber\\
&&+{\eta z_2^{-1}z_3^{-1}\over{\mu-z_2^{-1}z_3^{-1}}}\big({z_1^{-1}z_2^{-2}\over{\mu-\eta z_1^{-1}z_2^{-2}}}+{1\over{\eta-z_1^{-1}z_2^{-1}z_3^{-1}}}\big)+
{\eta z_1^{-1}z_2^{-1}z_3^{-1}\over{(\eta-z_1^{-1}z_2^{-1}z_3^{-1})(\eta-\mu z_1^{-1}z_2^{-2})}}\Big],
\label{eq:F31}
\end{eqnarray}
\begin{eqnarray}
F_{3,2}(\epsilon,k,q)&=&{16g^4\omega_0^4\over{(z_1-z_1^{-1})(z_2-z_2^{-1})^2(z_3-z_3^{-1})}}\Big[{z_1z_2z_3\over{\eta-z_1z_2z_3}}\big({z_1z_2\over{\mu- z_1z_2}}+
{\eta\over{\eta-\mu z_2^2z_3}}\big)+{\mu z_1z_2\over{(\mu-z_1z_2)(\mu-\eta z_2^2z_3)}}\nonumber\\
&&+{z_1^{-1}z_2^{-1}z_3^{-1}\over{\eta-z_1^{-1}z_2^{-1}z_3^{-1}}}\big({z_1^{-1}z_2^{-1}\over{\mu- z_1^{-1}z_2^{-1}}}+{\eta\over{\eta-\mu z_2^{-2}z_3^{-1}}}\big)+
{\mu z_1^{-1}z_2^{-1}\over{(\mu-z_1^{-1}z_2^{-1})(\mu-\eta z_2^{-2}z_3^{-1})}}\Big],
\label{eq:F32}
\end{eqnarray}
\begin{eqnarray}
F_{3,3}(\epsilon,k,q)&=&{16g^4\omega_0^4\over{(z_1-z_1^{-1})(z_2-z_2^{-1})^2(z_3-z_3^{-1})}}\big[{z_2z_3\over{\mu-z_2z_3}}-{z_2^{-1}z_3^{-1}\over{\mu-z_2^{-1}z_3^{-1}}}\big]
\big[ {z_1z_2\over{\mu-z_1z_2}}-{z_1^{-1}z_2^{-1}\over{\mu-z_1^{-1}z_2^{-1}}}\big].
\label{eq:F33}
\end{eqnarray}
\end{widetext}
Here $\mu=\exp{(iq)}$ and $\eta=\exp{(ik)}$.

\section{Appendix B}
\subsection{Derivation of the spectral function in the continuous limit.}
Eq.(\ref{nlinSch}) in 1D in the weak coupling limit $t>>E_p$ when the polaron radius is much larger than the lattice constant has an analytic solution. In that limit
Eq.(\ref{nlinSch}) may be rewritten in a differential form. Indeed, expanding $\psi_{\mathbf{n+m}}$ in a Taylor series up to the second order in the lattice constant
Eq.(\ref{nlinSch}) has the form:
\begin{equation}
-ta^2{d^2\psi(x)\over{dx^2}}-2E_p|\psi(x)|^2\psi(x)=(E+2t)\psi(x).
\label{nlinSch_diff}
\end{equation}
It is easy to check that the function
\begin{equation}
\psi^0(x)=({E_p\over{4t}})^{1/2}{1\over{\cosh{(E_p x/2at)}}}
\label{gr_State}
\end{equation}
is the solution of Eq.(\ref{nlinSch_diff}) corresponding to the bound state energy $E_0=-2t-E_p^2/4t$. Therefore, Eq.(\ref{shredinger}) with deformation field $\varphi^0(x)$
from Eq.(\ref{selfconsist}) may be rewritten as:
\begin{equation}
-ta^2{d^2\psi(x)\over{dx^2}}-{E_p^2\over{2t}}{\psi(x)\over{\cosh^2{(E_px/2at)}}}=(E+2t)\psi(x).
\label{Sch_deform}
\end{equation}
Rescaling the spatial variable $E_px/2at \to x$ this equation is reduced to the well known equation \cite{KabanovAlexandrov2008}:
\begin{equation}
-{d^2\psi(x)\over{dx^2}}-{2\psi(x)\over{\cosh^2{(x)}}}=\epsilon\psi(x).
\label{Sch_dimensionless}
\end{equation}
This equation is integrable and has analytic eigenfunctions. Therefore, the excited states of Eq.(\ref{Sch_deform}) have the following form:
\begin{equation}
\psi(x)=\Bigl ({2ikat\over{E_p}}-\tanh{({E_p x\over{2at}})}\Bigr )\exp{(ikx)}
\label{WF-excited}
\end{equation}
 with the excited state energies $E_k=-2t+tk^2a^2$. These wave functions satisfy the periodic boundary conditions when
 \begin{equation}
 ka\tan{(kL/2)}+{E_p\over{2t}}tanh{({E_pL\over{4ta}})}=0,
 \label{per_condit}
 \end{equation}
 where $L=Na$. This equation has $N-1$ roots for corresponding to $N-1$ itinerant states, because one state splits to localised polaron level $E_0$.
 These values of $k$ together with $\psi^0(x)$ (Eq.(\ref{gr_State})) define the complete set of eigenfunctions of Eq.(\ref{Sch_deform}).  The Fourier transform of these
 eigenfunctions defines the matrix elements Eqs.(\ref{Fk},\ref{Fkk}) which determine the polaron spectral function.

\end{document}